%% file: main.tex
\documentclass[letterpaper]{article}
\usepackage[preprint]{aaai2027}
\usepackage[hyphens]{url}
\usepackage{graphicx}
\usepackage{natbib}
\usepackage{caption}
\usepackage{booktabs}
\usepackage{mathtools}
\usepackage{amssymb}
\usepackage{multirow}
\usepackage{pifont}
\usepackage[table]{xcolor}
\usepackage{algorithm}
\usepackage{algpseudocode}
\usepackage[capitalize,noabbrev]{cleveref}

\graphicspath{{figures/}}

\providecommand{\Description}[1]{}

\title{From Chunks to Functional Evidence: Function-Aware Retrieval for EDA Documentation QA}
\input{authors}

\begin{document}
\maketitle
\input{sections/0abstract}
\input{sections/1introduction}
\input{sections/2background}
\input{sections/2method}
\input{sections/4experiment}
\input{sections/5conclusion}
\bibliography{references}
\clearpage
\appendix
\setcounter{table}{0}
\renewcommand{\thetable}{A\arabic{table}}
\input{sections/appendix}

\end{document}

%% file: authors.tex
\author{
Xiaotian Qiu\textsuperscript{1,2}, Kairui Liu\textsuperscript{1}, Shi Chenyi\textsuperscript{1},
Jinyuan Deng\textsuperscript{1}, Qi Sun\textsuperscript{1}, Cheng Zhuo\textsuperscript{1}
}
\affiliations{
\textsuperscript{1}Zhejiang University\\
\textsuperscript{2}Shanghai Innovation Institute
}

%% file: sections/0abstract.tex
\begin{abstract}
Retrieval-Augmented Generation (RAG) is widely used to ground answers in
documents. For complex technical documentation, however, the primary
bottleneck is often not model reasoning but a mismatch between a query and the
way knowledge is organized for retrieval. This mismatch is pronounced in
Electronic Design Automation (EDA) documentation, where the information
needed for an answer is scattered across heterogeneous yet tightly coupled
artifacts. We therefore redesign the basic retrieval unit of RAG. Instead of
operating on isolated chunks or binary relations, we collect typed artifacts
into EDA functional units. Each unit is recorded as a hyperedge with links to
its source chunks. We then train an encoder to align queries with functional
units and combine unit retrieval with direct chunk retrieval. After mapping
the selected units back to their sources, a unified reranker chooses the
evidence given to the generator. On the newly constructed EDADocEval-QA
dataset, our method improves ROUGE-L by 37.1\% over Chunk RAG and 55.6\% over
the strongest graph baseline. On the public ORD-MMBench benchmark, it improves
ROUGE-L by 30.0\% over the strongest baseline. These results support
function-aware evidence organization in the evaluated EDA documentation
settings.
\end{abstract}

%% file: sections/1introduction.tex
\section{Introduction}

\begin{figure}[!t]
  \centering
  \includegraphics[width=0.9\columnwidth]{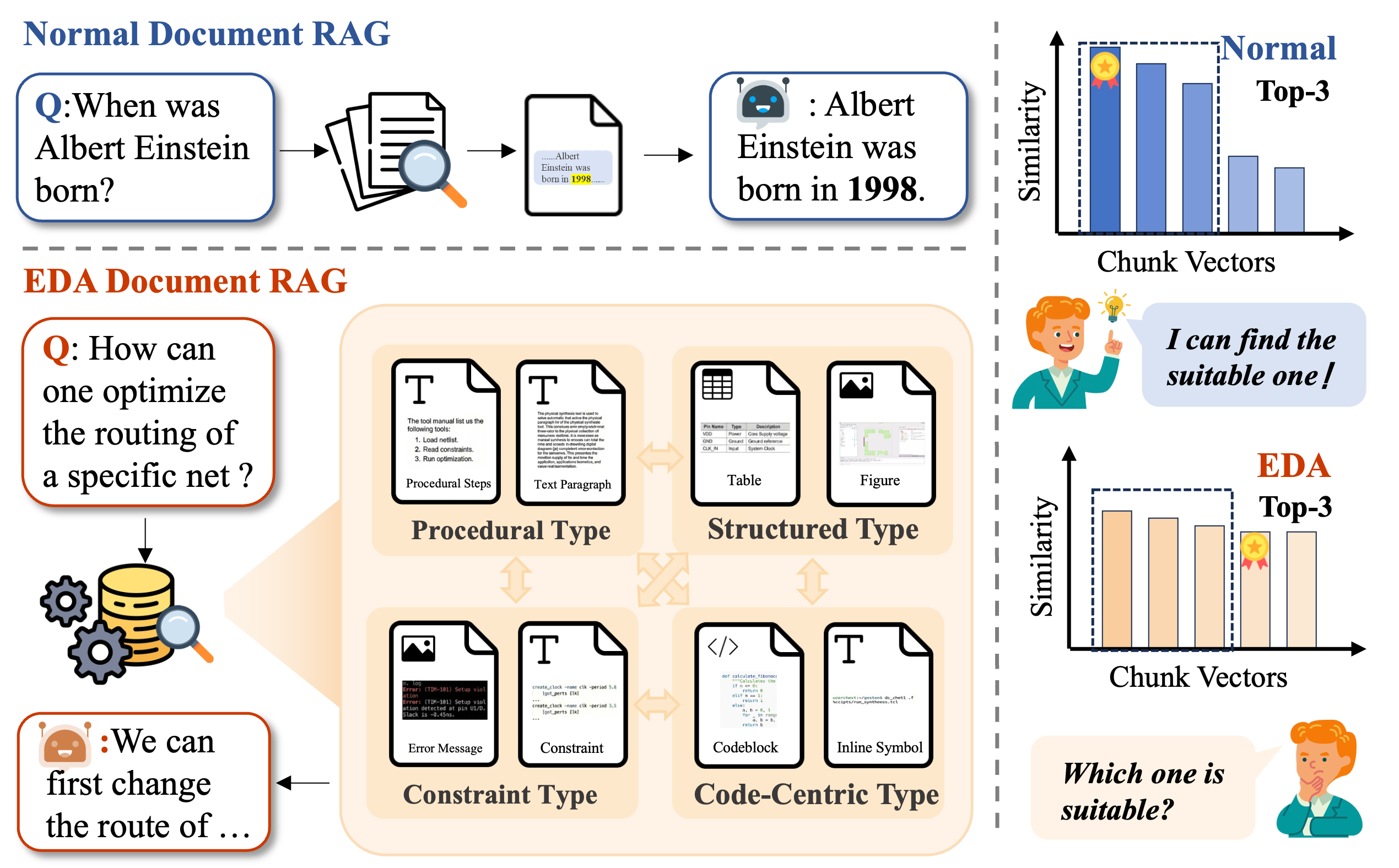}
  \caption{An illustrative embedding diagnostic: top-ranked EDA chunks have
  more tightly clustered similarity scores than the general-domain example,
  motivating retrieval units that preserve complementary functional roles.}
  \Description{Illustration comparing retrieval distinctiveness between EDA
  and general documents.}
  \label{fig:insight}
\end{figure}

Electronic Design Automation (EDA) tools support the complete circuit design
flow through tightly coupled implementation stages~\cite{ajayi2019openroad}.
Their documentation exemplifies a class of
technical sources with intrinsically complex knowledge organization.
Information is distributed across code blocks, parameter tables, figures,
procedures, constraints, and descriptive text that may appear in different
sections but jointly specify one tool function. We refer to this mixture of
heterogeneous evidence forms as \emph{multi-type} documentation. Answering a
practical question often requires composing several such artifacts rather than
retrieving one isolated passage, as illustrated in \Cref{fig:insight}.

RAG is a natural way to ground answers in these documents, but its primary
bottleneck here is often not model reasoning. It is the mismatch between a
question and the way evidence is organized for retrieval. Conventional dense
retrievers treat each chunk as an independent unit and assume that vector
similarity separates relevant from irrelevant passages. In EDA documentation,
topically similar chunks may instead describe different commands, flow stages,
or operating conditions. Figure~\ref{fig:insight} illustrates how repeated
domain terminology compresses the score differences among top-ranked chunks.
The retrievable unit should therefore preserve which artifacts participate in
the same function, as well as where each artifact came from.

This challenge calls for rethinking how knowledge is organized in
RAG~\cite{lewis2020rag}. Chunk RAG retrieves independently encoded
passages~\cite{karpukhin2020dense}. When dependent evidence appears in
different sections, this passage-level organization does not preserve their
functional grouping. GraphRAG organizes entities and community summaries for
global sensemaking~\cite{edge2024graphrag}, while HippoRAG uses graph
traversal for multi-hop retrieval~\cite{gutierrez2024hipporag}. Neither
necessarily retains the source context of a higher-order EDA usage event.
Hypergraphs provide a convenient representation in
which one unit can connect several artifacts~\cite{feng2019hypergraph}, but
the representation alone does not decide which artifacts belong together.
Generic name repetition or local co-occurrence may still connect evidence that
is topically similar yet functionally incompatible.

To address these issues, we redesign the basic retrieval unit around EDA
functionality. Our framework organizes typed artifacts into functional units,
uses a hypergraph to retain their higher-order membership and source
provenance, and retrieves the original evidence that supports a question.
Such grounded documentation QA also provides an evidence layer for EDA
assistants that must select commands, set parameters, and interpret warnings
before acting. Our contributions are summarized as follows:
\begin{itemize}
    \item We characterize the mismatch between conventional retrieval units
    and EDA documentation: artifacts needed for one answer are functionally
    coupled but can be separated by chunking or flattened by binary relations.
    
    \item We introduce EDA functional units as retrievable knowledge units.
    Occurrences remain local to their sources before they are grouped around a
    functional anchor. Hyperedges preserve their types and provenance and
    support conservative connections across documents.
    
    \item We propose query--unit contrastive alignment that distinguishes
    semantically similar units according to the artifact types available to
    answer a query.
    
    \item We develop a function-aware hybrid retriever that combines
    functional unit retrieval with direct chunk retrieval and returns a unified
    evidence context to the generator.
\end{itemize}

We evaluate our method against representative baselines on the public
ORD-MMBench benchmark~\cite{pu2025mm} and on EDADocEval-QA, a newly constructed
domain-specific dataset. EDADocEval-QA covers parameter lookup, syntax
retrieval, procedure synthesis, and constraint analysis. Our method obtains
0.473 ROUGE-L and 93.6 LLM-Score on this dataset, while the public benchmark
provides a complementary evaluation of generation and document retrieval.

%% file: sections/2background.tex
\section{Related Work}

\subsection{LLMs and RAG in EDA}

LLMs are widely used in EDA, demonstrating immense potential in tasks such as code generation~\cite{liu2023verilogeval, thakur2024verigen, lu2024rtllm}, EDA scripting~\cite{wu2024chateda}, CPU design and verification~\cite{wang2024chatcpu}, and conversational hardware design~\cite{blocklove2023chipchat, chang2023chipgpt}.
Leading industrial efforts, such as ChipNeMo~\cite{liu2023chipnemo}, have further validated the feasibility of domain-adaptive LLMs.
However, EDA documentation presents significant challenges for general LLMs due to proprietary protocols, complex parameter constraints, and specific terminologies.
LLMs often generate hallucinations or outdated information due to the lack of domain knowledge.
RAG addresses this issue by retrieving factual information from external databases~\cite{guu2020realm}.
Recent studies have optimized RAG for EDA by fine-tuning embedding models on domain corpora~\cite{pu2024rag_eda}.
Despite these improvements, most systems still organize evidence as independent chunks or generic relations rather than explicit EDA functional roles.

\subsection{Evidence Organization in RAG}

A key characteristic of EDA documents is the tight coupling of multi-type entities. 
Existing RAG paradigms make different trade-offs when handling this characteristic.

\textbf{Chunk-based RAG} treats documents as linear text sequences, utilizing sparse retrieval (e.g., BM25~\cite{robertson2009bm25}) or dense retrieval (e.g., DPR~\cite{karpukhin2020dense}) to fetch segments.
This approach can separate cross-entity dependencies.
A functional definition often appears in one section, while its timing constraints appear in another.
Chunk-based retrieval can therefore return isolated fragments.
Consequently, the LLM may receive an incomplete evidence chain~\cite{xing2026chipmind}.

\textbf{Graph-based RAG} extracts entities and relations to construct knowledge graphs~\cite{pan2024unifying}.
GraphRAG uses entity graphs and community summaries for global sensemaking~\cite{edge2024graphrag}, while HippoRAG uses knowledge graphs and Personalized PageRank for multi-hop retrieval~\cite{gutierrez2024hipporag}.
In the EDA setting studied here, however, entity- or relation-level retrieval may still separate a higher-order usage event in which code blocks, parameter tables, and procedural descriptions jointly support an answer.
Compressing detailed procedural descriptions or code blocks into relational summaries can also omit source context needed for generation.

\subsection{Structured Evidence Organization}

Recent retrieval methods increasingly organize evidence above the level of
independent chunks. SG-Filter builds a hierarchical summarized-semantic index
to distinguish otherwise similar passages~\cite{ye2025sg}. Evidence Units
groups visual assets with their contextual text through ontology-grounded
document organization~\cite{han2026evidence}, while TechDocRAG preserves
relations among clauses, tables, figures, and procedural steps in technical
documents~\cite{lee2026techdocrag}. Hypergraphs offer another representation:
a single hyperedge can connect multiple artifacts~\cite{feng2019hypergraph,
yadati2019hypergcn}, and recent systems use this property to organize
higher-order evidence for multi-hop retrieval~\cite{luo2025hypergraphrag,
zhang2026hkvm,an2026hyce}. Our framework uses this representation to retain
source-localized EDA occurrences, but defines retrieval units around tool
functions and learns to match queries with the artifact roles available in
each unit.

\begin{figure*}[!t]
  \centering
  \includegraphics[width=0.95\textwidth]{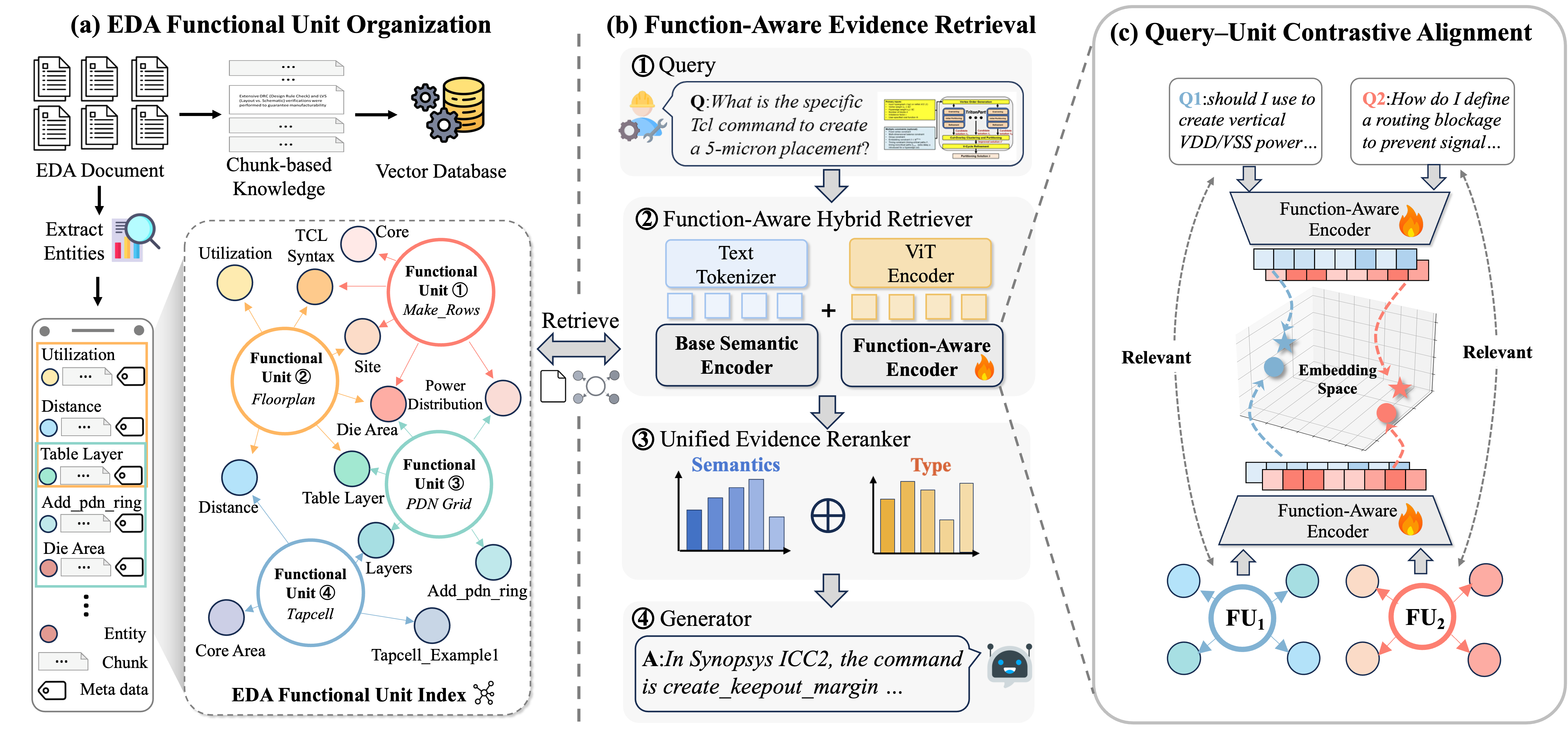}
  \caption{Framework overview. (a) Typed, source-localized EDA artifacts are organized into functional units, each represented as a hyperedge linked to its source chunks. (b) A function-aware hybrid retriever searches functional units and source chunks, after which a unified reranker selects the generation evidence. (c) Query--unit contrastive alignment trains the function-aware encoder.}
  \Description{The system organizes EDA artifacts into functional units, searches functional unit and source chunk indexes, reranks candidate evidence, and learns query--unit alignment.}
  \label{fig:overview}
\end{figure*}

\subsection{Reasoning-Augmented RAG}

Reasoning-enhanced frameworks such as Chain-of-Thought (CoT)~\cite{wei2022chain} and Tree of Thoughts (ToT)~\cite{yao2023tree} improve LLM reasoning by decomposing problems into intermediate steps.
IRCoT~\cite{trivedi2023interleaving} interleaves retrieval with reasoning, while ReAct~\cite{yao2023react} interleaves reasoning with external actions such as search.
Their efficacy still depends on the evidence retrieved at each step.
Using generic vector-based RAG or web search APIs proves ineffective for the specialized and proprietary nature of EDA documentation.
Our framework is complementary: it encodes typed documentation dependencies in
the retrieval index so that a reasoning method can begin from retrieved units
with intact source context.

%% file: sections/2method.tex
\section{Method}

\subsection{Problem Formulation and Overview}

We formulate EDA documentation QA as an evidence retrieval problem. Given a
query $q$ and document chunks $\mathcal{C}$, the system selects a generation
context $\mathcal{C}_q\subseteq\mathcal{C}$ that covers the facts and artifact
types needed for the answer. Fixed chunks often separate evidence that belongs
to the same EDA operation, while binary relations divide an event involving
several complementary roles. Before receiving a query, we organize the corpus
into source chunks and EDA functional units. A unit groups compatible
occurrences around one canonical function while retaining the source location
of every artifact. We represent it as a typed hyperedge linked to its source
chunks.

Our framework has three components. First, EDA functional unit organization
groups artifacts that play complementary roles in the same operation. Second,
query--unit contrastive alignment learns which artifact combinations distinguish
one function from nearby alternatives. Third, function-aware evidence
retrieval searches the unit and chunk indexes in parallel.
Once selected for a query, the returned units and chunks become candidate
evidence. Selected units are expanded to their source chunks, and a unified
reranker chooses the top-$K$ chunks as $\mathcal{C}_q$.

\subsection{EDA Functional Unit Organization}
\label{sec:hypergraph}

We first identify typed artifacts and organize them into functional units
before any query is received. The resulting index is encoded as a hypergraph
$\mathcal{H}=(\mathcal{V},\mathcal{E}_{\text{hyper}})$: a node represents a
documentation artifact, and a hyperedge records the artifacts and source
chunks that constitute one unit. The hypergraph is therefore the
representation of the index; the substantive construction decision is which
artifacts are allowed to form a functional unit.

\paragraph{Occurrences within a source.}
Construction begins from document structure rather than a whole-corpus lexical
graph. A deterministic Markdown parser records headings, block boundaries,
line spans, and image references. Blocks under the same section are grouped
locally, with code blocks, ordered procedures, and diagnostic messages serving
as natural boundaries between distinct usage events. Very short neighboring
segments are merged only within the same document. This produces occurrence
candidates whose provenance is known before semantic extraction.

Each candidate is then decomposed into typed artifacts. Our schema
$\mathcal{T}$ covers code, inline commands, tables, figures, procedures,
descriptive text, constraints, diagnostics, and file artifacts. Deterministic
rules identify explicit structures such as code blocks and tables, while an
LLM identifies semantic spans such as procedures and constraints. For every
artifact $v$, we retain its content, artifact type
$\tau(v)$, source document, source chunk, and extraction metadata.

Artifact type and functional role are related but not interchangeable.
$\tau(v)$ describes the evidence form, such as \texttt{table} or
\texttt{codeblock}. A functional role describes how that artifact participates
in a particular usage event, for example as the operation, parameter
specification, condition, example, or diagnostic consequence. We infer this
relation from the surrounding source context and its functional anchor; it is
not a second supervised node label. This distinction matters later because the
training objective operates on observable artifact types, not on latent role
names.

Formally, a \emph{functional occurrence} from source region $r$ is
\begin{equation}
  o_r=(a_r,\mathcal{V}_r,\mathcal{C}_r,p_r),
\end{equation}
where $a_r$ is a command or file anchor, $\mathcal{V}_r$ contains the typed
artifacts in that local event, $\mathcal{C}_r$ contains its source chunks, and
$p_r$ stores document and line provenance. This definition is stronger
than local word co-occurrence: the grouped artifacts must belong to one
structural source region and remain traceable to it.

\paragraph{Functional units across sources.}
An occurrence within one source already forms a useful unit. Some functions,
however, are described across manuals: one document may define a command while
another provides a parameter table or worked example. We connect such
occurrences only through an admissible canonical anchor. Let
$\mathcal{O}(a)=\{o_r:\operatorname{norm}(a_r)=a\}$. A unit spanning documents is
created only when $\mathcal{O}(a)$ contains at least two source occurrences
from different documents and their union contains at least two artifact types.
The linker accepts stable command and file anchors, filters generic
placeholders and language type aliases, and caps the number of occurrences
attached to one anchor. These restrictions prevent frequent but irrelevant
strings from becoming hubs across the corpus.

Each retained unit is represented as a hyperedge
$e\in\mathcal{E}_{\text{hyper}}$:
\begin{equation}
  e=(\mathcal{V}(e),S(e),\operatorname{chunks}(e),P(e)),
\end{equation}
where $\mathcal{V}(e)\subseteq\mathcal{V}$ is the artifact set, $S(e)$ is its
retrieval summary, $\operatorname{chunks}(e)\subseteq\mathcal{C}$ is the
deduplicated set of original chunks, and $P(e)$ retains the contributing
occurrences and source references. We serialize $S(e)$ by artifact type so
that the unit embedding retains
$\mathcal{T}(e)=\{\tau(v)\mid v\in\mathcal{V}(e)\}$. The summary is used for
unit retrieval only; the generator receives the original source chunks reached
through $\operatorname{chunks}(e)$.

\paragraph{Validation and updates.}
Structural checks reject empty units, missing source references, invalid type
labels, and links without sufficient type diversity. They establish
structural validity and provenance, not the semantic accuracy of LLM
extraction. Construction can resume from its cache, so validated outputs survive an
interrupted call. When a document is added or changed, the parser and extractor
run only on its chunks. The linker then resolves their anchors against the
retained index and recomputes the affected units; a deletion follows the same
provenance links in reverse. Consequently, adding the 33rd document does not
require extracting the unchanged 32-document corpus again.

\paragraph{Concrete example.}
Consider the \texttt{define\_pdn\_grid} command in OpenROAD.
Its default \texttt{-starts\_with: GROUND}, syntax, and strap-orientation constraints occur in different source regions.
The canonical command anchor connects the parameter table, code block, and
constraint paragraph, while their source IDs remain intact.
The resulting functional unit is represented by a hyperedge with type
composition
$\{\texttt{table},\texttt{codeblock},\texttt{text\_paragraph}\}$ and
backtracks to all three chunks.

\subsection{Query--Unit Contrastive Alignment}
\label{sec:contrastive}

Effective retrieval requires more than locating semantically similar text. It
must also identify functional units whose available artifact types can support
the query.
Since EDA documents tend to cluster more tightly in the embedding space, relying solely on semantic similarity often fails to distinguish among candidates, resulting in recalls that appear semantically relevant but lack the required evidence forms.
Our contrastive objective aligns queries with units that contain the required
combination of types while preserving general semantic similarity.

We use the structured summary $S(e)$ as the textual interface for encoding each
functional unit. For units that include visual or structured entities, we
serialize them into the multimodal input format so that the unit embedding
reflects both semantic content and artifact type. This shifts supervision from
local chunk matching to query--unit alignment.

Training queries are constructed from functional units, so each query $q$ and
its reference unit set $\mathcal{P}_q$ share the target type combination by
construction. We learn the alignment through contrastive positive and negative
pairs together with an artifact coverage ranking objective. For each query, we
select one unit from $\mathcal{P}_q$ as the positive sample $e^+$ and sample
$\mathcal{N}(q)$ from the remaining units. Hard negatives are semantically
plausible but lack a critical artifact type; random negatives preserve general
semantic discrimination.

Hard negatives capture a common failure mode in multi-type documentation:
candidates can be topically related yet miss evidence needed for an answer.
For instance, a purely descriptive paragraph may be semantically relevant to a parameter query but lacks the concrete table row that specifies defaults or ranges.
By contrasting such candidates against the positive unit, the model learns to
prefer usable type combinations within a narrow semantic neighborhood. We use
a mixed sampler: units with partial overlap provide difficult type negatives, while
semantically hard or random units keep the learned space from collapsing into
a type classifier. The query is always textual, whereas a candidate unit may
combine textual and visual content. This asymmetry teaches the query
representation to select the evidence forms available in a unit without
requiring a separate type router at inference.

We use pretrained Visualized-BGE-M3 as the \emph{Base Semantic Encoder}.
Applying Low-Rank Adaptation (LoRA) with the objective below produces the
\emph{Function-Aware Encoder}. Both encoders accommodate textual and visual
entities in multimodal EDA documents.
The semantic component follows the InfoNCE formulation:
\begin{equation}
    \mathcal{L}_{\mathrm{NCE}}(q) = -\log \frac{\exp(s^+ / \tau)}{\exp(s^+ / \tau) + \sum_{e^- \in \mathcal{N}} \exp(s^- / \tau)}
\end{equation}
where $s^+=\mathrm{sim}(q,e^+)$, $s^-=\mathrm{sim}(q,e^-)$, and $\tau$ is the temperature.
Let $\mathcal{T}_q$ denote the artifact types required by training query $q$
and $\mathcal{T}(e)$ the types present in candidate unit $e$. We define
$\kappa_q(e)=|\mathcal{T}_q\cap\mathcal{T}(e)|/|\mathcal{T}_q|$, with
$\kappa_q(e)=1$ when $\mathcal{T}_q$ is empty. For the positive and sampled
negative candidates, let $s_i=\mathrm{sim}(q,e_i)/\tau$ and
$\mathcal{R}_q=\{(i,j):i<j,\ |\kappa_q(e_i)-\kappa_q(e_j)|>0.01\}$.
For each $(i,j)\in\mathcal{R}_q$, write
$\Delta^\kappa_{ij}=\kappa_q(e_i)-\kappa_q(e_j)$ and
$\Delta^s_{ij}=s_i-s_j$. A pairwise margin term then orders candidates with
different artifact coverage:
\begin{equation}
 \mathcal{L}_{\mathrm{type}}=\frac{1}{|\mathcal{R}_q|}\!\sum_{(i,j)\in\mathcal{R}_q}\!
 \max\!\left(0,m-\operatorname{sgn}(\Delta^\kappa_{ij})\Delta^s_{ij}\right).
\end{equation}
We set $\mathcal{L}_{\mathrm{type}}=0$ when $\mathcal{R}_q$ is empty and use
margin $m=0.1$; the final objective is
$\mathcal{L}=\mathcal{L}_{\mathrm{NCE}}+\lambda\mathcal{L}_{\mathrm{type}}$.
At inference time, the model does not explicitly predict a type set. Instead,
the learned embedding space recalls functional units whose artifact composition
matches the query.

\subsection{Function-Aware Evidence Retrieval}
\label{sec:retrieval}

The function-aware hybrid retriever uses two complementary indices. The
unit index captures the evidence forms associated with an EDA operation. The
chunk index preserves access to localized details. Searching both is
important: functional units provide broader coverage, whereas direct chunk
retrieval can recover a precise detail that is weakly represented in a unit
summary.

Given $q$, we retrieve the top-$k_p$ functional units using the Function-Aware Encoder
and expand each unit through $\text{chunks}(e)$. Expansion is limited to the
source chunks that belong to the selected units; it does not traverse unrelated
neighbors. In parallel, the Base Semantic Encoder retrieves the top-$k_c$
source chunks. At this point, the retrieved units and chunks become candidate
evidence. We merge and deduplicate their source chunks to obtain
$\mathcal{C}_{\text{cand}}$.

The unit path is not an unrestricted graph walk. The explicit
$\operatorname{chunks}(e)$ mapping produces a candidate set conditioned on the
function retrieved for the query. This prevents a unit with a high score from
pulling in unrelated neighbors that merely share a generic entity. The chunk
path plays the complementary role: it can recover a local command option,
table row, or warning that is highly relevant even when its unit summary is
broad. Their union separates two decisions that a single retriever would
conflate: locating the relevant function and locating the precise source span
within or near that function.

The unified evidence reranker scores every candidate in both encoder spaces:
\begin{equation}
    s(c)=\alpha\,\mathrm{sim}(q_{\mathrm{base}},c_{\mathrm{base}})
    +(1-\alpha)\,\mathrm{sim}(q_{\mathrm{func}},c_{\mathrm{func}})
\end{equation}
where $\alpha=0.4$ and each query--chunk similarity is computed in one
internally aligned encoder space.
We select the top-$K$ chunks as the generation context $\mathcal{C}_q$.

The Function-Aware Encoder carries the type preference learned in
Section~\ref{sec:contrastive}, while the Base Semantic Encoder preserves local
semantic matching. Their fusion determines the source chunks seen by the
generator rather than using unit retrieval only as an auxiliary
signal. Because both paths terminate at source chunks, generation is grounded
in the same evidence format used by conventional RAG. Functional units change
how candidate evidence is organized and selected, not what is presented to the
generator.

%% file: sections/4experiment.tex
\section{Experiments}

\subsection{Experimental Setup}

\paragraph{Evaluation settings.}
We use two complementary EDA documentation QA settings.
(1) \textbf{EDADocEval-QA} contains 100 QA pairs over 33 OpenROAD technical documents (324K tokens), with 25 questions in each of four categories.
The four query types have distinct evidence compositions: Parameter Lookup (PL), Syntax Retrieval (SR), Procedure Synthesis (PS), and Debug \& Constraint (DC).
These categories require different combinations of structured and textual evidence.
(2) \textbf{ORD-MMBench}~\cite{pu2025mm} is a public multi-modal EDA benchmark with text and screenshot queries over EDA documentation; we report overall generation quality and official document-retrieval performance.
ORD-MMBench provides an independent public evaluation within the broader EDA documentation setting.

\paragraph{Baselines.}
We consider three families of baselines that represent mainstream RAG paradigms.
\textbf{Chunk-level RAG} includes Chunk RAG~\cite{lewis2020rag}, which performs dense retrieval over fixed-length document chunks.
\textbf{Graph-based RAG} includes GraphRAG~\cite{edge2024graphrag}, which organizes evidence with an entity-relation graph and community-based summarization, LightRAG~\cite{guo2024lightrag}, which targets lightweight graph maintenance, HippoRAG 2~\cite{gutierrez2025hipporag2}, which incorporates memory-augmented retrieval, and HyperGraphRAG~\cite{luo2025hypergraphrag}, which extends graph retrieval to multi-ary relations.
\textbf{Reasoning-augmented frameworks} include ReAct~\cite{yao2023react}, which interleaves reasoning with tool use, and Search-o1~\cite{li2025searcho1}, which augments reasoning with agentic search.
For ORD-MMBench, we use the benchmark's reported baselines:
VisRAG~\cite{yu2024visrag}, EchoSight~\cite{yan2024echosight},
RAG-EDA~\cite{pu2024rag_eda}, and VISTA-RAG~\cite{zhou2024vista},
which span text-based and multimodal retrieval.

\paragraph{Implementation.}
We use Qwen2.5-72B as the LLM backbone for answer generation.
The retrieval encoder is Visualized-BGE-M3 (873M parameters) fine-tuned with LoRA (rank=8, $\alpha$=16, dropout=0.1) on a single NVIDIA RTX 4090 GPU, yielding 786K trainable parameters (0.09\% of total).
Training data consists of 1,250 query--unit pairs automatically generated from 250 training packs, with an 85/15 train/validation split.
The contrastive objective combines InfoNCE with a type-aware margin ranking loss weighted by $\lambda{=}0.3$ (temperature $\tau{=}0.07$, 7 negatives per query using mixed hard-negative mining).
Training runs for 10 epochs with AdamW (lr=$10^{-4}$) and early stopping; the best checkpoint is selected at epoch 8 (val loss = 0.735).
For functional-unit retrieval, we set top-$M$=10; for chunk retrieval, top-$N$=5; the final context uses top-$K$=5 after unified reranking.
Unless a row is explicitly labeled DeepSeek-R1, EDADocEval-QA systems use Qwen2.5-72B for answer generation; all systems use the same 33-document corpus and QA/evaluation protocol, while retaining their method-specific evidence organization.
For ORD-MMBench, we use GPT-4o as the generation backbone to align with the original benchmark configuration.

\paragraph{Metrics.}
We report BLEU~\cite{papineni2002bleu}, ROUGE-L~\cite{lin2004rouge}, and a 0--100 LLM-Score that sums five 0--20 ratings for correctness, completeness, relevance, professionalism, and conciseness against the reference answer.
Because Qwen2.5-72B is both a generator and the LLM-Score judge for several systems, this metric is complementary rather than independent factuality evidence; we therefore interpret it together with lexical and retrieval metrics.

\subsection{Main Results}

\begin{table*}[!t]
  \centering
  \small
  \setlength{\tabcolsep}{10pt}
  \begin{tabular}{lccc}
    \toprule
    Method & ROUGE-L $\uparrow$ & BLEU $\uparrow$ & LLM-Score $\uparrow$ \\
    \midrule
    \multicolumn{4}{l}{\textbf{Chunk-level RAG}} \\
    \midrule
    Chunk RAG (Qwen2.5-72B)~\cite{lewis2020rag} & 0.345 & 0.052 & 88.3 \\
    Chunk RAG (DeepSeek-R1)~\cite{lewis2020rag} & 0.327 & 0.097 & 87.5 \\
    \midrule
    \multicolumn{4}{l}{\textbf{Graph-based RAG}} \\
    \midrule
    GraphRAG~\cite{edge2024graphrag} & 0.226 & 0.050 & 87.8 \\
    LightRAG~\cite{guo2024lightrag} & 0.219 & 0.024 & 92.8 \\
    HyperGraphRAG~\cite{luo2025hypergraphrag} & 0.304 & 0.094 & 77.2 \\
    HippoRAG 2~\cite{gutierrez2025hipporag2} & 0.301 & 0.168 & 88.0 \\
    \midrule
    \multicolumn{4}{l}{\textbf{Reasoning framework}} \\
    \midrule
    ReAct~\cite{yao2023react} & 0.374 & 0.091 & 84.7 \\
    Search-o1~\cite{li2025searcho1} & 0.283 & 0.044 & 89.9 \\
    \midrule
    \textbf{Ours} & \textbf{0.473} & \textbf{0.189} & \textbf{93.6} \\
    \bottomrule
  \end{tabular}
  \caption{Main results on EDADocEval-QA. LLM-Score is a five-part quality
  score (0--100) evaluated by Qwen2.5-72B. Best results are in bold.}
  \label{tab:main_results}
\end{table*}

Table~\ref{tab:main_results} reports results on EDADocEval-QA under a unified RAG setting.
We next analyze component contributions and construction cost before evaluating
generation and retrieval on the public ORD-MMBench benchmark.

As shown in Table~\ref{tab:main_results}, our method achieves ROUGE-L 0.473, BLEU 0.189, and LLM-Score 93.6 on EDADocEval-QA, ranking first on all three metrics.
Compared with Chunk RAG using Qwen2.5-72B, it improves ROUGE-L by 0.128 (0.345$\rightarrow$0.473), BLEU by 0.137 (0.052$\rightarrow$0.189), and LLM-Score by 5.3 points (88.3$\rightarrow$93.6).
The simultaneous increase in lexical metrics and LLM-Score is consistent with improved alignment to answer-critical details, although the LLM-based score is not independent factuality evidence.

Reasoning-augmented methods provide a strong baseline: ReAct achieves ROUGE-L 0.374, outperforming most retrieval-only baselines, yet still trails our method by 0.099 in ROUGE-L and 0.098 in BLEU.
This pattern is consistent with evidence organization remaining important when
support is fragmented across heterogeneous artifact types; our functional units
explicitly target completeness across evidence types.

We also observe metric divergence where some methods obtain high LLM-Score but low lexical metrics.
For example, LightRAG reaches LLM-Score 92.8 but only 0.219 ROUGE-L and 0.024 BLEU.
The metrics therefore capture different properties of the generated answers;
without answer-level analysis, their disagreement does not identify a single
cause. Our method improves lexical metrics together with LLM-Score, which is
consistent with retrieving both complementary evidence and answer details.

\subsection{Ablation Study}

\begin{table}[!t]
  \centering
  \small
  \setlength{\tabcolsep}{5pt}
  \begin{tabular}{lcc}
    \toprule
    Configuration & ROUGE-L & $\Delta$ \\
    \midrule
    Chunk RAG & 0.345 & -- \\
    + Functional unit organization & 0.426 & +23\% \\
    + Function-aware hybrid retrieval & 0.461 & +34\% \\
    + Query--unit alignment & \textbf{0.473} & \textbf{+37\%} \\
    \bottomrule
  \end{tabular}
  \caption{Cumulative ablation of the three method components.}
  \label{tab:ablation}
\end{table}

We ablate functional unit organization, hybrid retrieval, and query--unit
alignment on EDADocEval-QA.
As shown in Table~\ref{tab:ablation}, organizing the corpus into functional
units provides the largest gain, improving ROUGE-L from 0.345 to 0.426 (an
increase of 0.081, or 23\%). This indicates that grouping artifacts by their
functional roles mitigates evidence fragmentation.
This is consistent with EDA queries requiring parameters, code snippets, and procedural descriptions that participate in the same function but occur in different document regions.

The function-aware hybrid retriever further raises ROUGE-L to 0.461, an
increase of 0.035 over functional unit retrieval alone. This result indicates
that source chunks complement functional units in this setup. Query--unit
alignment adds 0.012, reaching 0.473. Since this final increment is small and
no confidence interval is available, we do not interpret it as a statistically
established standalone gain.

\subsection{Construction and Retrieval Cost}

We measured hypergraph construction on the full 33-document corpus using
Qwen3.6-27B. The pipeline processed 404 chunks and extracted 3,650 typed nodes.
Every chunk retained its source location and cached metadata, none required
fallback processing, and all structural consistency checks passed. We did not
manually label extraction accuracy, so this result verifies index completeness
and internal consistency rather than semantic correctness. Building the index
required 427,824 LLM tokens and 1,668.0 seconds, with peak RSS of 84.0 MiB.
At query time, the function-aware hybrid retriever reduced the average generation
context from 1,697 to 1,415 tokens, a reduction of 16.6\%, while retrieval
latency increased from 20.69 to 27.67 ms.

\paragraph{Incremental updates and parser robustness.}
To test whether the index can absorb new material without rebuilding the full
corpus, we first constructed it from 32 documents and then added the remaining
document. The updater reused 402 existing units and rebuilt two. All seven
semantic outputs matched those produced by rebuilding the index from all 33
documents. We also applied four formatting changes to the source Markdown,
including alternative code fences and repeated blank lines. Every variant
produced the same semantic graph, containing 404 hyperedges and 2,481 canonical
nodes, covered all 33 documents, and retained no invalid units. These
experiments test index maintenance and parser stability, not the semantic
precision or recall of extracted entities.

\subsection{ORD-MMBench Results}

We evaluate on the public ORD-MMBench benchmark using GPT-4o as the generation backbone to align with the original configuration.

\begin{table}[!t]
  \centering
  \small
  \setlength{\tabcolsep}{5pt}
  \begin{tabular}{lccc}
    \toprule
    Method & LLM-Score & BLEU & ROUGE-L \\
    \midrule
    VisRAG & 20.4 & 0.084 & 0.225 \\
    EchoSight & 57.9 & 0.139 & 0.276 \\
    RAG-EDA & 49.1 & 0.134 & 0.268 \\
    VISTA-RAG & 69.6 & 0.180 & 0.317 \\
    \textbf{Ours} & \textbf{79.6} & \textbf{0.240} & \textbf{0.412} \\
    \bottomrule
  \end{tabular}
  \caption{Overall ORD-MMBench generation results. Baseline values are from
  MM-GRADE~\cite{pu2025mm}.}
  \label{tab:ord_mmbench_results}
\end{table}

As shown in \Cref{tab:ord_mmbench_results}, our method achieves the best overall performance, with an LLM-Score of 79.6, BLEU of 0.240, and ROUGE-L of 0.412.
Compared with the strongest baseline VISTA-RAG, it improves overall ROUGE-L by 0.095 (0.317$\rightarrow$0.412), BLEU by 0.060 (0.180$\rightarrow$0.240), and LLM-Score by 10.0 points (69.6$\rightarrow$79.6).

Document retrieval reaches Recall@1/5 of 0.700/0.917 versus 0.679/0.896 for MM-GRADE, providing an independent public retrieval comparison alongside the generation results.

ORD-MMBench assigns one target document to each query, so its official
retrieval metric is document-level Recall@K.
We compare BGE-M3~\cite{chen2024m3embedding},
RAG-EDA~\cite{pu2024rag_eda}, DEDR~\cite{salemi2023dedr},
BLIP~\cite{li2022blip}, VisRAG~\cite{yu2024visrag},
EchoSight~\cite{yan2024echosight}, Visualized-BGE-M3~\cite{zhou2024vista},
and MM-GRADE~\cite{pu2025mm}.

\begin{table}[t]
  \centering
  \small
  \setlength{\tabcolsep}{1.5pt}
  \begin{tabular}{lccccc}
    \toprule
    Method & R@1 & R@2 & R@3 & R@4 & R@5 \\
    \midrule
    BGE-M3 & 0.590 & 0.694 & 0.731 & 0.761 & 0.776 \\
    RAG-EDA & 0.530 & 0.612 & 0.679 & 0.724 & 0.739 \\
    DEDR & 0.112 & 0.187 & 0.231 & 0.284 & 0.299 \\
    BLIP & 0.127 & 0.209 & 0.276 & 0.299 & 0.313 \\
    VisRAG & 0.025 & 0.042 & 0.050 & 0.083 & 0.108 \\
    EchoSight & 0.383 & 0.492 & 0.583 & 0.625 & 0.667 \\
    Visualized-BGE-M3 & 0.582 & 0.716 & 0.739 & 0.761 & 0.799 \\
    MM-GRADE & 0.679 & 0.828 & 0.843 & 0.866 & 0.896 \\
    \midrule
    \textbf{Ours} & \textbf{0.700} & \textbf{0.850} & \textbf{0.858} & \textbf{0.883} & \textbf{0.917} \\
    \bottomrule
  \end{tabular}
  \caption{Document retrieval on ORD-MMBench. Baseline values are from
  MM-GRADE~\cite{pu2025mm}.}
  \label{tab:ord_mmbench_retrieval}
\end{table}

As detailed in \Cref{tab:ord_mmbench_retrieval}, the retrieval advantage is
consistent but modest at every cutoff: the margin over MM-GRADE ranges from
0.015 to 0.022. Text-oriented retrievers remain substantially stronger than
vision-only retrieval on this benchmark. These results support evidence
coverage as one contributor to the generation gains, rather than attributing
the full difference to the hypergraph alone.

\subsection{Discussion}

Table~\ref{tab:main_results} compares complete systems under a shared corpus,
question set, and evaluation protocol; the DeepSeek-R1 row is the only
explicitly labeled generator change. Since retriever adaptation differs across
baseline families, this table measures end-to-end performance rather than the
isolated effect of hypergraph structure. The cumulative ablation offers a more
direct comparison within our framework, while ORD-MMBench tests transfer to a
public EDA benchmark. These results do not establish generality beyond
technical EDA documentation.

The method also introduces a practical trade-off. Retrieving functional units
alongside source chunks reduces the context sent to the generator but adds
retrieval latency. Direct chunk retrieval remains useful when one local passage
is sufficient or construction misses an occurrence. The reported results
measure answer quality, document retrieval, construction consistency, and
runtime cost, but not the semantic accuracy of intermediate extraction. Each
table entry comes from one completed evaluation; we report neither multi-seed
averages nor statistical significance.

%% file: sections/5conclusion.tex
\section{Conclusion}

We presented a framework that organizes typed EDA artifacts into functional
units and uses query--unit alignment with hybrid retrieval to select evidence
from functional units and source chunks. It improves EDADocEval-QA ROUGE-L by
37\% over Chunk RAG and transfers to ORD-MMBench. The study remains limited to
one domain-specific dataset and one public EDA benchmark; construction requires
offline LLM extraction, and hybrid retrieval reduces generation context at the
cost of latency. Future work should evaluate extraction accuracy and additional
EDA tool ecosystems.

%% file: sections/appendix.tex
\section{Indexing and Retrieval Procedure}

\begin{algorithm}[H]
  \caption{Functional-unit indexing and function-aware retrieval}
  \label{alg:functional_retrieval}
  \footnotesize
  \begin{algorithmic}[1]
    \Require documents $\mathcal{D}$, query $q$, trained encoders
      $f_{\mathrm{base}},f_{\mathrm{func}}$, budgets $k_p,k_c,K$; $\alpha=0.4$
    \Ensure generation context $\mathcal{C}_q$
    \Statex \textbf{Indexing}
    \State $\mathcal{O}\gets\emptyset$;
      $\mathcal{C}\gets\textsc{Chunks}(\mathcal{D})$
    \ForAll{$r\in\textsc{Parse}(d),\,d\in\mathcal{D}$}
      \State $\mathcal{V}_r\gets\textsc{ExtractTypedArtifacts}(r)$;
        $a_r\gets\textsc{CanonicalAnchor}(r,\mathcal{V}_r)$
      \State append $(a_r,\mathcal{V}_r,\textsc{Chunks}(r),
        \textsc{Prov}(r))$ to $\mathcal{O}$
    \EndFor
    \State $\mathcal{E}\gets\textsc{MergeAdmissibleOccurrences}(\mathcal{O})$
    \State store $S(e)$ and $\mathrm{chunks}(e)$ for $e\in\mathcal{E}$
    \Statex \textbf{Retrieval}
    \State $\mathcal{E}_q\gets
      \textsc{TopItems}_{k_p}(\mathcal{E};\,
      e\mapsto\mathrm{sim}(f_{\mathrm{func}}(q),f_{\mathrm{func}}(S(e))))$
    \State $\mathcal{C}_u\gets
      \bigcup_{e\in\mathcal{E}_q}\mathrm{chunks}(e)$
    \State $\mathcal{C}_c\gets
      \textsc{TopItems}_{k_c}(\mathcal{C};\,
      c\mapsto\mathrm{sim}(f_{\mathrm{base}}(q),f_{\mathrm{base}}(c)))$
    \State $\mathcal{C}_{\mathrm{cand}}\gets
      \textsc{Deduplicate}(\mathcal{C}_u\cup\mathcal{C}_c)$
    \State $s(c)\gets\alpha\,\mathrm{sim}(f_{\mathrm{base}}(q),f_{\mathrm{base}}(c))
      +(1-\alpha)\,\mathrm{sim}(f_{\mathrm{func}}(q),f_{\mathrm{func}}(c))$
    \State \Return $\mathcal{C}_q\gets
      \textsc{TopItems}_{K}(\mathcal{C}_{\mathrm{cand}};\,c\mapsto s(c))$
  \end{algorithmic}
\end{algorithm}

\section{Incremental Indexing and Parser Invariance}

\subsection{Incremental Updates}

Starting from a 32-document index, we inserted \texttt{27Flute3.md} and
compared the update with a fresh 33-document build. Only the added document and
anchors affected by it were recomputed.

\noindent\begin{minipage}{\columnwidth}
  \centering
  \small
  \setlength{\tabcolsep}{3pt}
  \begin{tabular}{@{}lr@{}}
    \toprule
    Measurement & Result \\
    \midrule
    Unchanged functional units reused & 402 \\
    Functional units built from new document & 2 \\
    Index outputs matching full build & 7/7 \\
    Incremental index assembly & 0.407 s \\
    Full index assembly & 1.396 s \\
    \bottomrule
  \end{tabular}
  \captionof{table}{Incremental update from 32 to 33 documents. The seven
  matched files are the hyperedge, node, provenance, bipartite-edge,
  chunk-content, type-bag, and rejection records; single-run times are shown.}
  \label{tab:supp_incremental_audit}
\end{minipage}

\subsection{Markdown Normalization Invariance}

We independently indexed four equivalent corpus variants: closing heading
hashes, repeated blank lines, tilde code fences, and a UTF-8 byte-order mark.

\noindent\begin{minipage}{\columnwidth}
  \centering
  \small
  \setlength{\tabcolsep}{3pt}
  \begin{tabular}{@{}lr@{}}
    \toprule
    Measurement & Result \\
    \midrule
    Variants with identical semantic graph & 4/4 \\
    Variants with byte-identical serialized output & 3/4 \\
    Functional-unit hyperedges per variant & 404 \\
    Source documents covered & 33/33 \\
    Invalid retained units per variant & 0 \\
    \bottomrule
  \end{tabular}
  \captionof{table}{Index invariance under equivalent Markdown syntax.
  Semantic-graph equality is determined by hash.}
  \label{tab:supp_markdown_audit}
\end{minipage}

All four variants preserve the semantic graph. Repeated blank lines change the
serialized-output hash while preserving the semantic-graph hash.

\section{Qualitative Example}

{\small
\noindent\textbf{Query.} How does the
\texttt{-sink\_clustering\_enable} parameter affect the clock tree synthesis
process in the \texttt{clock\_tree\_synthesis} command?\par
\noindent\textbf{Reference content.} Pre-clustering creates one sub-tree level
before the H-tree; each cluster buffer becomes an H-tree endpoint.\par
\noindent\textbf{Chunk RAG.} ``\ldots enables sink clustering during the clock
tree synthesis process, which groups clock sinks into clusters to optimize the
clock tree structure.''\par
\noindent\textbf{Ours.} ``\ldots enables pre-clustering of sinks to create one
level of sub-tree before building the H-tree, where each cluster is driven by a
buffer that becomes an endpoint of the H-tree structure.''\par
\noindent One functional-unit-derived and four directly retrieved chunks recover
the pre-H-tree subtree and buffer-endpoint relation.
}